\documentclass[a4paper,12pt] {article}
\usepackage{jpc}
\usepackage{lscape}
\usepackage{graphicx}
\usepackage{subfigure}
\usepackage{amssymb}

\begin{document}
\title{Non-equilibrium dynamics in amorphous Si$_3$B$_3$N$_7$ }

\author{
A.\ Hannemann, J.\ C.\ Sch\"on, M.\ Jansen
\\
{Max-Planck-Institut f\"ur Festk\"orperforschung } \\
{Heisenbergstr. 1, D-70569 Stuttgart, Germany} \\
P. Sibani \\
Fysisk Institut - SDU, Campusvej 55, DK-5230 Odense, Denmark}
\maketitle
\begin{abstract}
\small
We present extensive numerical investigations of the
structural relaxation dynamics
of a realistic model of  the amorphous
high-temperature ceramic a-Si$_3$B$_3$N$_7$, probing
  the mean square displacement (MSD) of the atoms,
 the bond survival probability (BSP), the average energy,
the specific heat, and the two-point energy average.
Combining the information from
these different sources, we identify   a transition
temperature $T_c \approx 2000$ K below which the
system is no longer ergodic and
physical quantities observed over a time $t_{obs}$ show  a systematic
parametric dependence  on the waiting time  $t_w$, or age,  elapsed after
the quench. The aging dynamics 'stiffens' as the system
becomes older, which is similar to the behavior of highly
idealized models such as Ising spin-glasses and Lennard-Jones glasses.
\end{abstract}
Pacs.No 02.50.-r,61.43.Fs,64.70.Pf,65.60.+a
\normalsize
\newpage

\section{Introduction}
Physical aging of polymers and other 'soft' materials
    was studied experimentally  almost three decades   ago  by
Struik and co-workers~\cite{VanTurnhout77,Struik78}.
The  interest  of a wider   community for this area was  then greatly  stimulated
by work of   Lundgren et al.~\cite{Lundgren83} on the magnetic properties of
spin glasses, and  aging  has since  remained  a  source  of  important  questions
and insights in non-equilibrium statistical physics~\cite{Young98}.

The  basic   features of aging were mainly  established
during   the eighties in  studies of magnetic linear susceptibility and   autocorrelation function
of spin glasses ~\cite{Alba86,Nordblad86,Alba87,Svedlindh87},  but are
shared by  many other systems,  including, to cite only a few examples,
type II superconductors~\cite{Nicodemi01}, glasses~\cite{Kob00a,Utz00,Barrat99a,Barrat01a,Berthier02},
 granular materials~\cite{Josserand00},
and soft condensed matter~\cite{Cipelletti00}.
Following a rapid quench of a  relevant parameter, e.g.\ when the temperature of a  glass is rapidly
  lowered below a threshold value, aging materials
are unable to  re-equilibrate  within the available
observation time. What is  observed is a  slow  change of   physical properties
which occurs at a  decelerating pace, and in a fashion   largely independent of
most details of the  microscopic interactions. More precisely,   physical  observables
systematically depend on the system age   $t_w$, i.e.\ the time  elapsed from the  initial
quench, and
the observation time $t_{obs}$. Based on the  relative values of $t_w$ and $t_{obs}$, two 
dynamical  regimes can be identified: 
 For     $t_{obs} \ll t_w $,  macroscopic physical averages are  approximately constant,
 and two-time correlations  are related to  the linear response through  
 the  fluctuation-dissipation theorem. 
Such   a 'quasi-equilibrium' regime 
 strongly resembles  the  time translational invariant  dynamics of thermal equilibrium, 
except for  the fact   that
many  quantities, e.g. relaxation times,  can carry  a parametric age  dependence.
For     $t_{obs} \gg t_w $,  physical averages visibly    drift  and
two-time averages no longer   obey the fluctuation dissipation theorem.

Recent improvements in time resolved spectroscopic techniques~\cite{Bissig03,Cipelletti03a,Buisson03,Buisson03a},  
have highlighted that, far from being a smooth flow,  the drift of aging systems 
 results from  a highly  irregular, so-called  \emph{intermittent},
process, whereby   sudden and  large
configurational re-arrangements~\cite{Sibani03} 
punctuate the prevailing  equilibrium-like fluctuations.
These intermittent events may correspond    to irreversible jumps from one metastable
state to the next~\cite{Sibani04,Crisanti04}.

Aging has raised a wealth of theoretical questions,
for example,  whether  the non-equilibrium 
aging regime can be
described through the concept of effective temperature~\cite{Cugliandolo97,Calabrese04}.
Of  direct  relevance for the present investigations
is the origin of the so-called pure aging behavior, i.e. 
the  $t_{obs}/t_w$ scaling of various observables
in the aging regime~\cite{Sibani03,Bouchaud92} and of  
modifyied scalings~\cite{Rinn01a} which are called sub-aging  and 
super- or even hyper-aging.   

 This  work  discusses numerical investigations
 of the dynamical and thermodynamical properties
 of a \emph{realistic} model 
of a-Si$_3$B$_3$N$_7$ for a large temperature range.
The material is  a representative of 
a class of nitridic
ceramics with chemical 
composition 
a-Si$_x$B$_y$N$_1$C$_z$\cite{Jansen02a,Seyferth90a,Baldus92a,Su93a,Funayama93a,Jansen97a,Baldus99a,Aldinger98a,Srivastava98a,Riedel96a}
which can  be   synthesized via the sol-gel process\cite{Jansen97a}.
These ceramics are of great technological interest and are presently  considered 
for high-temperature engine applications.

Nitridic ceramics appear to form amorphous covalent networks with a homogeneous
distribution of the cations at least down to a scale of
1 nm\cite{Wullen00b}. Experimental studies have  mainly focused on the quaternary
compound  a-SiBN$_3$C.  Under standard conditions,  both  a-SiBN$_3$C and
a-Si$_3$B$_3$N$_7$ are  thermally
stable and amorphous up to very high temperatures, e.g.\ about $2100$ K
for a-SiBN$_3$C. They  possess  excellent
 mechanical and elastic properties: e.g.\ a-SiBN$_3$C
has a high bulk modulus of ca.\ $200 - 300$ GPa, and is stable  against oxidation up to $1700$ K.

The   amorphous structure suggests a
similarity with   glassy materials, and hence the presence
of  aging properties  of potential
technological relevance.
Probing   the  glass  transition of e.g.\ a-Si$_3$B$_3$N$_7$,
the  basic representative of this class of nitridic ceramics,
is   impossible  at
standard conditions, since decomposition takes place  at $T\approx 1900$ K,
i.e.\  well before the ceramic melts. However, the melting transition might
be accessible, as we have argued, at very high pressures~\cite{Hannemann03d}.

The    MC simulations presented below
aim at  the determination   of the  temperature range below
which the non-ergodic behavior of a-Si$_3$B$_3$N$_7$ sets in, and
at the  characterization of the aging dynamics in that low temperature regime.
Admittedly, determining "for all practical  purposes" whether
a system is ergodic or not  is not straightforward and is possibly
an ill-posed question\cite{Vazquez03a}. As will become clear
from the discussion, we refer to  non-ergodic behavior as
a collection of dynamical properties usually associated with glassiness,
which, taken together, describe a strong deviation from thermal equilibrium.

To detect the onset of ergodicity breaking and
characterize the behavior in both the
ergodic and non-ergodic regions, we  consider  a number of relevant
dynamical and thermodynamical measures:
1) The physical movement of the particles, as statistically
characterized by the Mean Square Displacement (MSD),
is diffusive in the ergodic region, and  sub-diffusive elsewhere.
2) The related Bond Survival Probability (BSP)\cite{Kob99b} drops to a finite
average close to zero on a characteristic time scale at
sufficiently high temperatures but becomes scale-free
and age dependent in the non-ergodic region. There,
it  decreases as a logarithmic  function
of $t_{obs}/t_w^{\alpha}$, where $\alpha$ is close to one
and  depends  on the type of   particles forming the bond.
3) The specific heat $C_V$ is calculated in
three  different ways, which  agree in equilibrium but
markedly differ in the non-ergodic region.
4) The two-time energy-energy average $\phi(t_w, t_{obs};T)$
defined in equation \ref{phidef}. By construction, $\phi$ equals one in equilibrium or in a  quasi-equilibrium regime,
and is larger than one when the system is out of equilibrium.
5) The age dependence of the  energies of the local minima
or so-called inherent structures (IS)\cite{Stillinger82a} characterizing the trajectories
in the non-ergodic regime. These energies decrease  logarithmically with the age.

The rest of the paper is structured as follows: in Section~\ref{Section_model}
we introduce the model and simulation technique. In  Section~\ref{Section_ergodicity}
we define the  statistical measures used  to probe the ergodicity breaking and
present the corresponding results. Finally, Section~\ref{Section_discussion}  contains a broader discussion
of the results, where we show that our findings are in general agreement with previous studies
of the aging dynamics of much simpler   atomistic models~\cite{Barrat99a,Sastry99a,Crisanti00a}
and also with experimental results in e.g.\ spin-glasses~\cite{Nordblad97a}, polymer systems\cite{Struik78}
and charge-density waves~\cite{Biljakovic91a}.

\section{Model and Techniques}
\label{Section_model}
The model of a-Si$_3$B$_3$N$_7$ consisted of $162$ Si-atoms, $162$ B-atoms and
$378$ N-atoms, respectively, in a $19.1 \times 19.1 \times 19.1$ {\AA}$^3$
cubic box. As an interaction potential, we employed a two-body potential from
the literature\cite{Gastreich00b} based on fits to ab-initio energy
calculations of crystalline and molecular compounds
containing Si-, B- and N-atoms. This potential successfully
reproduces experimental data regarding
 structural properties such as bond lengths,  and bulk
 properties, e.g.\ vibrational frequencies, of the binary compounds
Si$_3$N$_4$ and  BN and of molecules containing
Si$-$N$-$B units.

The starting configurations for our  simulations were
generated by relaxation from high-temperature melts.
The simulations were performed at  fixed temperature and volume, with a
Monte-Carlo algorithm using
the Metropolis acceptance criterion.
In each update,  an atom is randomly selected for
an attempted move with  a random direction,  and with an average  size
 chosen to achieve  an acceptance rate
of $\approx 50$ \%. One Monte-Carlo
cycle  (MCC) corresponds  to $N_{atom} =702$ such  individual
moves.
Note  that  the kinetic energy ($3/2 k_B T$ per atom) does not
appear in  MC-simulations, and that all   quantities studied
relate to the configurational energy. For comparison,
we also performed MD simulations in the melt. Comparing
mean square displacements determined from MD and MC simulations,
respectively, we estimate that (at least in the high-temperature regime)
 one MCC corresponds to about 0.5 fs.
At  low temperature, the value of the microscopic time scale
is not known precisely. However, this value is not crucial
for the study of aging quantities  which, as in our case,
have a logarithmic time dependence or depend on the ratio of two
 times\cite{Huitema99a}.

The temperatures investigated ranged from 25  to
10000 K; however, for $T \le 250$ K no discernible dynamical evolution
 took place on the time scale of the simulations.
For each temperature up to $3000$ K and above $3000$ K, $9$ and $3$
runs, respectively, of length $t_{total} = 2 \times 10^5$  MCC  were performed.

The energy $E(t;T)$ as function of time  was
registered every $10$ MCC. Along the individual trajectories for $T =
250,...,7000$ K, halting points $x_H$ were chosen, from which
conjugate gradient minimizations were performed.
In the following,  $t_{init} \approx 1000$ MCC is the initialization
time of the MC-simulations needed for the system to reach equilibrium in the
ergodic regime (i.e.\ at high temperatures), while $t_w \ge t_{init}$ is the waiting
time before the observations  begin. Furthermore, an ensemble of
$100$ runs of length $t_{total} = 10^6$ MCC was studied, in order
 to investigate aging in greater detail, and with better statistics.
 Each starting configuration of the ensemble was generated by quenching the system from $4000$ K
 to $1250$ K.
 \footnote{Due to the rather expensive energy evaluation, performing the runs in this ensemble
  required approximately a computing time of  one year on a AMD/1800 PC.}

 The diffusivity of the atoms in the high temperature
 regime was calculated straightforwardly from the mean square
 displacement as a function of time via the Einstein-Smochulowski relation
 ($D \propto \frac{MSD(t_{obs})}{t_{obs}}$).
 Finally, in the  bond survival probability data, a
 B-N or Si-N bond was considered  broken if the
 interatomic distance was above $2.0$ {\AA}.

In computing time averages and correlations  we considered the
observation interval $I=[t_w,t_w+t_{obs}]$. Time averages over $I$
are denoted by an over-bar, while an angular  bracket  stands
for an average over an ensemble of independent trajectories.


\section{Ergodicity-breaking and aging behavior}
\label{Section_ergodicity}
\subsection{Ergodicity breaking}

Evidence for ergodicity breaking can be extracted from
a change of behavior in dynamical quantities i.e.\  the MSD and BSP, as well as from
thermodynamical quantities, i.e.\  the heat capacity, the energy average and the
two-point energy average. All these data concur   that ergodicity breaking occurs
 but differ to some degree with respect to the
precise value of the transition temperature
$T_c$. The latter can be approximately located  within the range $2000-2500$ K.

We consider the high temperature behavior first.
For $T>T'=2100$ K, the MSD between pairs of time-displaced atom positions
 \begin{equation}
MSD(t_{obs}, t_w) = \sum_i \left[ \vec{r}_i(t_w+t_{obs}) - \vec{r}_i(t_w) \right]^2.
\label{msd}
\end{equation}
only depends on the length of the observation interval $t_{obs} = (t_{obs} + t_w) - t_w $.
The diffusion coefficients for the different atoms all have
 an Arrhenius temperature dependence, with  activation barriers
$\Delta E= 1.24 \pm  0.09$ eV, $1.27 \pm 0.06$ eV for silicon and boron atoms, respectively.

\begin{figure}
\centering
\includegraphics[width=0.66\columnwidth,angle=-90]{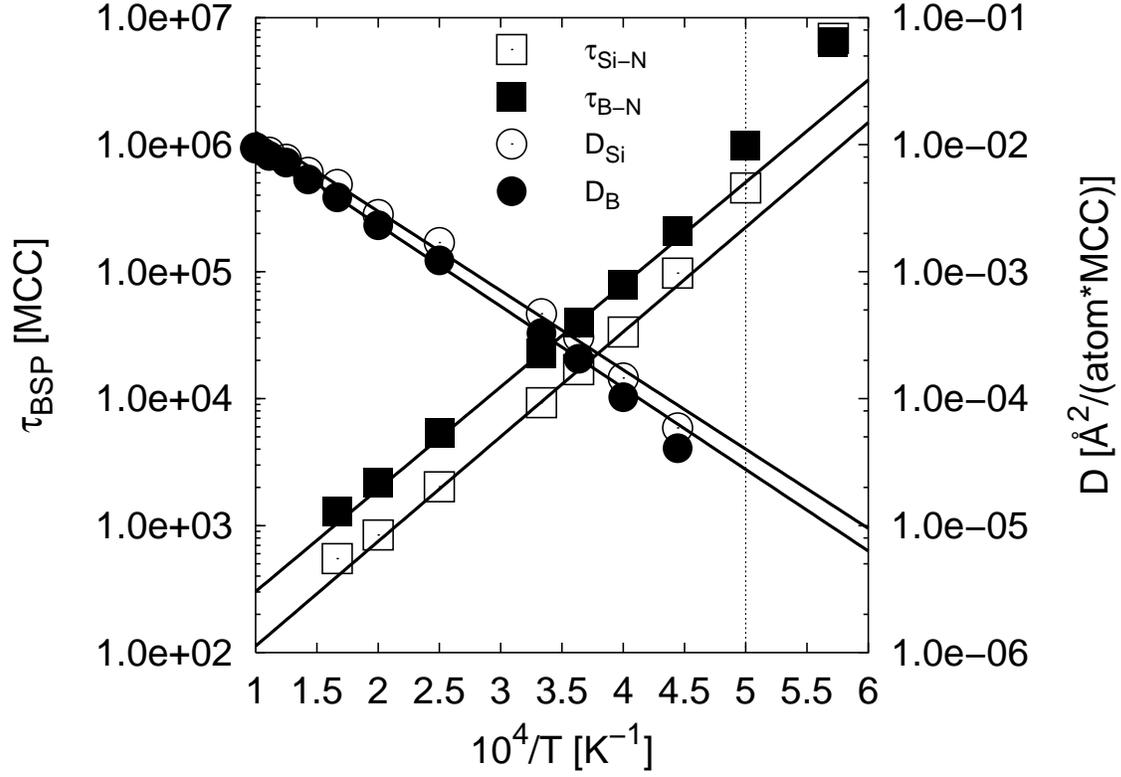}
\caption{
Temperature dependence of relaxation times $\tau_{BSP}$ for Si-N (empty squares) and B-N (filled squares), and
temperature dependence of diffusion coefficient $D(Si)$ (empty circles) and $D(B)$ (filled circles). Straight lines show
 fits to an Arrhenius law. The dotted vertical line indicates the critical temperature $T_C \approx 2000$ K.
\label{diffusionandbspinone}
}
\end{figure}

In the same regime, the  BSP is age independent and has a finite
  relaxation time $\tau_{BSP}$.
The  values of  $\tau_{BSP}$ were   obtained by fitting the time dependence of the
bond survival probability after the initialization time $t_{init}$
to  a
Kohlrausch-Williams-Watts law:\cite{Elliott90a}
\begin{equation}
BSP(t,t_{init};T) \propto \exp\left( -\left[\frac{t}{\tau_{BSP}(T)}\right]^{\beta}\right),
\end{equation}
yielding exponents $\beta \approx 0.3 - 0.5$.
The  time scale $\tau_{BSP}(T)$  has an Arrhenius-like temperature dependence, with
activation barriers $ \Delta E= 1.64 \pm  0.02$ eV and $1.60 \pm 0.02$ eV for
Si-N and B-N bonds, respectively. Arrhenius plots for
the diffusion coefficients of silicon and boron atoms
and for the relaxation times $\tau_{BSP}$ are plotted
in Fig.~\ref{diffusionandbspinone}. A deviation from the
Arrhenius scaling is apparent in both quantities  as $T$ approaches 2000 K from above.
The activation barriers found for  both diffusion and bond breaking are in a similar range,
suggesting that a single Arrhenius time scale might characterize the high temperature dynamics. However, although diffusive processes need to be accompanied by bond-breaking, it is not clear whether this effect or steric blocking by neighbor atoms dominates the diffusion at various temperature scales.

Deviations from activated behavior appear in the relaxation time as
the temperature decreases. Approaching  $T' \approx 2100$ K from above,
the diffusion coefficients  for  B-, Si- and N-atoms all vanish for $T \rightarrow T'$
as  $D \propto(T-T')^{\gamma}$ with $\gamma = 1.6$. A similar deviation of $\tau_{BSP}$
from an Arrhenius law  is observed  when  $T^{''}\approx 1720$ K for B-N
and $T^{''} \approx 1820$ K for Si-N, respectively, are  approached from above.
Close to these temperatures, $\tau_{BSP}(T)$ grows roughly like a power law:
$\tau_{BSP}(T) \propto (T/T^{''} -1)^{\delta}$
with   $\delta_{B-N} \approx -2.5$, and $\delta_{Si-N} \approx -2.3$.
At even lower temperatures, the particle motion becomes sub-diffusive and
the  BSPs fall off logarithmically.  Both quantities become scale free but
gain a characteristic age dependence, discussed in more detail
in section \ref{Subsection_aging_properties}.

Thermodynamical variables such as  the specific heat $C_V$,
the energy average, and a suitably normalized
two-time energy-energy average $\phi$ require
averaging over an ensemble. For a fixed amount of
numerical effort, they  are hence  less precisely
determined than 'local' dynamical properties, but offer
nevertheless concurring insights into ergodicity
breaking and aging behavior. 

The heat capacity was evaluated with three
different prescriptions, all agreeing in thermal equilibrium.
Note, that all specific heat data are reported on a per-atom basis.
The disagreement observed at lower  temperature  shows  that
the system is non-ergodic for the observational time scale used.
We note that agreement at higher temperatures does not neccessarily imply that the
equilibrium state is reached~\cite{Yu04}, since such an agreement is only a neccessary
but not a sufficient criterion for the system to be in equilibrium.

Our  first prescription  is naively  carried over from equilibrium thermodynamics.
Leaving the $T$ dependence of the mean energy understood on the
r.h.s. of the equation, we define
\begin{equation}
C_V^a(T)    =   \frac{\partial \langle \overline{E} \rangle}{\partial T} \label{eq2},
\end{equation}
where the temperature derivative is performed numerically after the  ensemble
and time  averaging denoted by angular brackets ($\langle\rangle$) and an overbar, respectively.

Secondly, we consider simulations  where the temperature is increased or
decreased by a small amount $\pm \Delta T \approx \pm  0.1 T$ at age $t_w$.
The trajectories are  then  time-averaged over the time $t_{obs} \ll t_w$
spent at $T \pm \Delta T$.

\begin{figure}
\includegraphics[width=0.53\columnwidth,angle=-90]{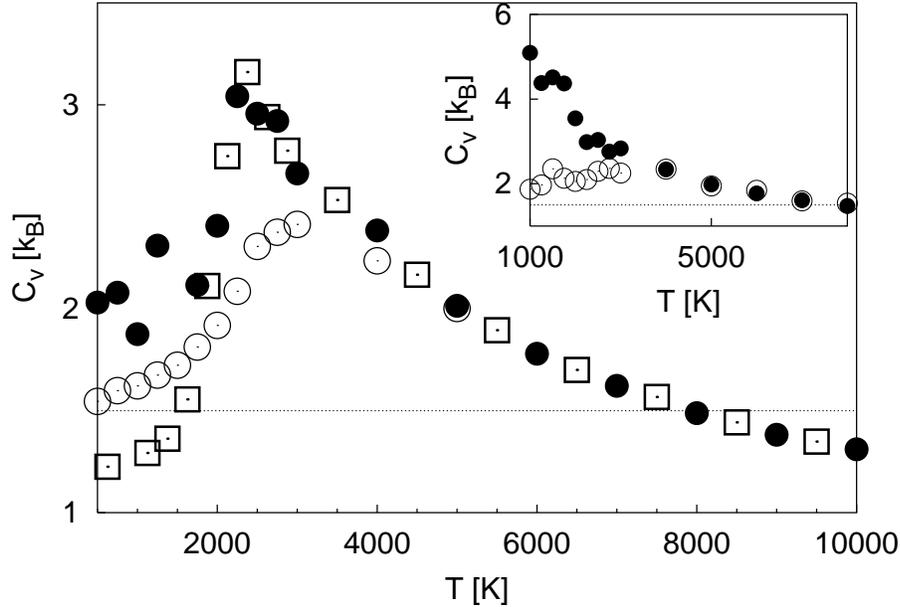}
\caption{Temperature dependence of the specific heats per atom $C_V^{a}$ ($\Box$),
$C_V^{b}$ ($\circ$) and $C_V^c$ ($\bullet$),  for $T \in [500,10000]$ K.
For $C_V^a$   and $C_V^c$, the
waiting time $t_w$ and the observation time $t_{obs}$
are  both equal to $10^5$ MCC. $C_V^b$ is
obtained by averaging over  $10$ equidistant  $t_w$  values
in the interval  $10^5 - 2 \cdot 10^5$, for each of three independent trajectories.
The  observation time was kept at  $t_{obs}=5\cdot10^3 \ll t_w$.
 All procedures agree for $T>3000$K, the region in which the system   equilibrates
according to this measure.
\emph{Inset}: For temperatures $T \in [1000,8000]$ K and waiting
time $t_w = 10^4$, $C_V^c$ is shown
with symbols $\circ$  and $\bullet$ corresponding to
observation times  $t_{obs}= 10^4$  and  $t_{obs}= 10^5$ respectively.
The high  values ($\bullet$)   at  low temperatures seen
  for $t_{obs} \gg t_w$ are a consequence of the
energy drift present in  this dynamical regime.
For the shorter observation time
($\circ$), the values  are in good agreement with
the $C_V^b$ values shown in the main panel with the same symbol (note the difference in scale).
The agreement expresses the equivalence between drift and fluctuation
within the quasi-equilibrium low temperature  regime of the aging system.
\label{fig_cv}
}
\end{figure}

This procedure yields
\begin{equation}
C_V^b (T) = \frac{\overline{E(T+\Delta T;t_w)} - \overline{E(T-\Delta T;t_w)} }{2 \Delta T}. \label{eq3}
\end{equation}
Finally,  we gauge the variance of the energy fluctuations within the observation interval
 $I=[t_w,t_w+t_{obs}]$ and estimate the heat capacity as
 \begin{equation}
C_V^c (T) =  \frac{\langle \overline{E^2}  -  \overline{E}^2 \rangle }{k_BT^2}\label{eq4},
\end{equation}
 for a range of observation times $t_{obs}$ which  straddles
 $t_w$.
Figure~\ref{fig_cv}, where all $C_V$ values are on a per-atom basis, shows that  all three expressions agree for temperatures above $3000$ K, but clearly
disagree below $2000$ K, which leads to a provisional value for the
ergodicity breaking temperature $T_c \approx 2500$ K.

Within the non-ergodic region $T< T_c$, we have studied $C_V^a(T)$ for $t_{obs} \ll t_w$,
   the quasi-equilibrium regime,   and for $t_{obs} \gg t_w$
 the non-equilibrium regime.  The dependence  of $C_V^a(T)$
 on $T$, $t_{obs}$ and $t_w$ can also be obtained from the fitted
time and age dependence of the mean energy given in  Eq.~\ref{energy_drift}
below,  by performing the time average and then differentiating with respect to
$T$.
For  $T < T_c$ and $t_{obs} \gg t_w$, we find that  $C_V^a$ has a small dependence on the length of the time interval $t_{obs}$ over which the time average is performed. This is
connected to  the logarithmic  drift of the energy, which moreover
is more pronounced  just below $T_c$ than  near  $T=0$  (c.f.\ fig.\ \ref{fig3}).

$C_V^b$ mimics an experiment performed after some relatively long equilibration time $t_w \gg t_{obs}$
and thus  yields the most "realistic" value for the specific heat for all temperatures.
$C_V^c$ provides a  way to check the  validity of the  fluctuation-dissipation theorem:
it   should coincide with    the   two other expressions  above $T_c$,
and additionally, below $T_c$  in the restricted quasi-equilibrium
region $t_{obs}<t_w$. Indeed, when  increasing
$t_{obs}$ past $t_w$  the observed dynamics in $C_V^c$ changes from quasi-equilibrium to off-equilibrium
in the temperature range between $2000$ K and $3000$ K (c.f.\ inset in fig.\ \ref{fig_cv}).
Recall that the (trivial) contribution to $C_V$ due to the change in kinetic energy is never
 included, i.e., in equilibrium $C_V \rightarrow 1.5 k_B$
 for $T \rightarrow 0$ K. Furthermore, the interaction potential $V(r)$
  vanishes for  $r \rightarrow \infty$,
and thus this contribution to $C_V$ goes to zero as $T \rightarrow \infty$.

As shown in fig.~\ref{fig_cv}, the  above  prescriptions for the specific heat
yield, as expected,  almost identical results in the   high-$T$
ergodic dynamical regime, but differ at low  $T$. This indicates that ergodicity is broken
 below $T_c \approx 2000 - 3000$ K.
 Even in the low temperature region, $C_V^b$ and $C_V^c$ nevertheless roughly coincide
 in a restricted time interval which
 corresponds to the quasi-equilibrium regime  $t_{obs} < t_w$. However,
 the inset clearly shows the large difference between $C_V^c$
 in the quasi-equilibrium and the non-equilibrium regime, for temperatures below $T_C$.

Yet another  probe of ergodicity-breaking  is  the
 two-time energy average defined by
\begin{equation}
\label{phidef}
\phi({t_w},t_{obs};T) = \frac{\langle E(t_w)E(t_w +
t_{obs})\rangle}{\langle E(t_w)E(t_w)\rangle}  \label{eq1},
\end{equation}
Above $T_c$, and for  $t_{obs}<t_w$  below $T_c$, we expect
the dynamics to be  time translational invariant, and hence $\phi=1$.  Considering that the
energy is negative and decreases monotonically with age,
we expect  an increase of $\phi$ with observation time, once
the dynamics gains access to states of considerably lower
energy.

\begin{figure}
\includegraphics[height=0.8\textheight]{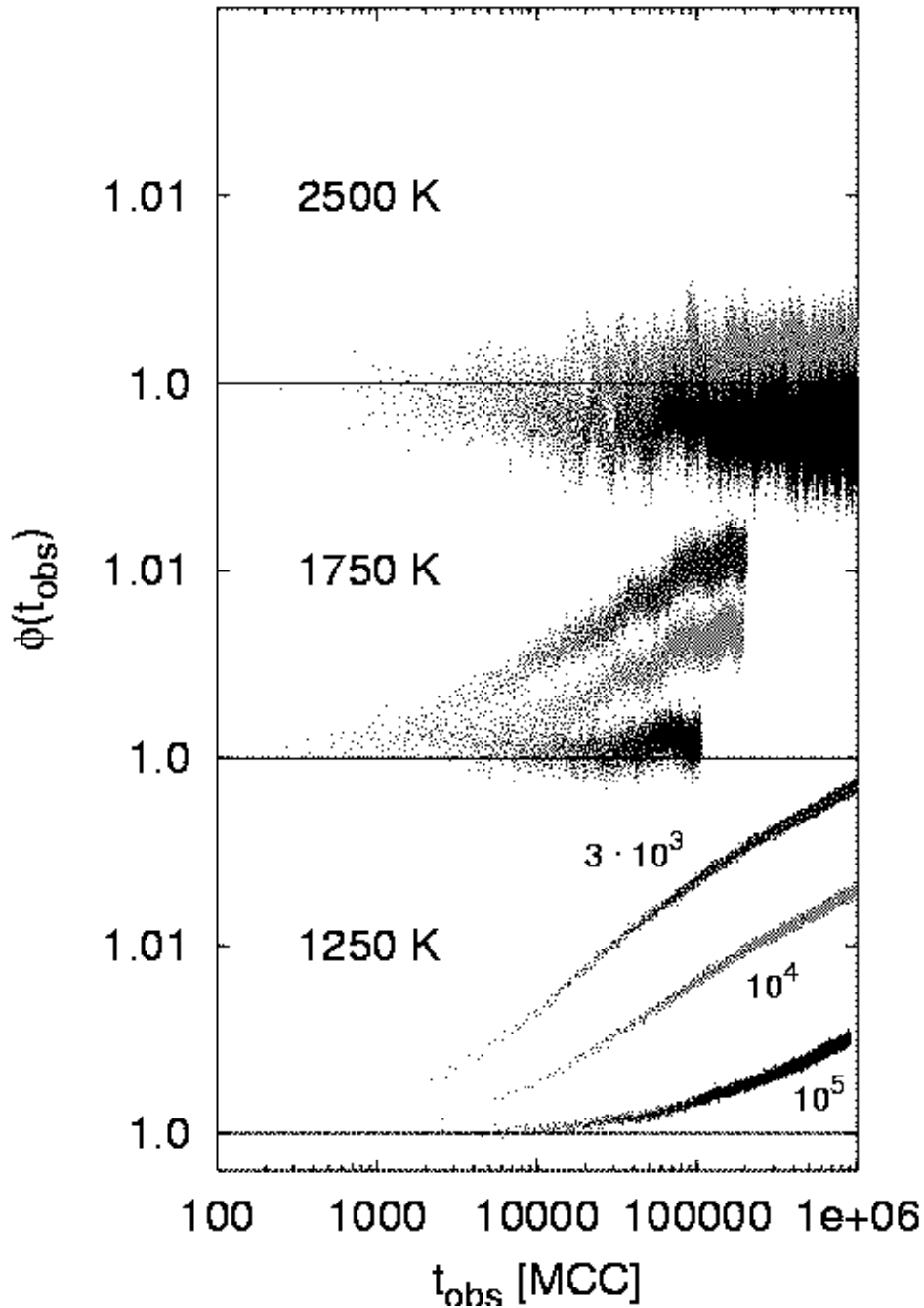}
\caption{
Energy self correlation functions $\phi(t_{obs})$ for three different
temperatures 1250 K, 1750 K and 2500 K (bottom to top).
Red, green and blue curves were recorded after waiting times
$t_w=3\cdot10^3$, $10^4$ and $10^5$ MCC, respectively.
The ensemble averages were calculated over 5 (1750 K), 6 (2500 K), and 100 (1250 K) different
trajectories. Number inside the lower
figure indicate waiting times $t_w$ (in MCC) after which measurements of
$\phi(t_{obs})$ began.
\label{fig_phi}
}
\end{figure}

Figure \ref{fig_phi} depicts $\phi$ for three temperatures,
$T = 1250$ K, $T = 1750$ K and $T = 2500$ K, and for three
values of the system age, $t_w = 3\cdot 10^3, 10^4$ and $10^5$
MCC.  The comparison clearly shows that ergodicity is broken as described at the lowest temperature (T =
 1250 K): due to the decrease of the configurational energy during aging, the two-time energy average
 increases as the observation time increases. At the intermediate temperature (T = 1750 K)
the aging effect is still visible but difficult to quantify, and at the highest temperature (T = 2500 K)
it has vanished completely.

Taking all theses indicators,
MSD, BSP, $C_V^{a,b,c}$ and $\phi$ into account,
we conclude that a-Si$_3$B$_3$N$_7$ exhibits ergodicity breaking at $T_c \approx 2000$ K.

\subsection{Aging properties}
\label{Subsection_aging_properties}
In order to describe  the non-ergodic regime in more detail
we have studied the age dependence of:  (\emph{i})  the energy of
the inherent structures and of  the mean potential  energy,
(\emph{ii}) the bond survival probability (BSP), (\emph{iii})
 the mean square displacement (MSD) of the particles, and
(\emph{iv}) the two point correlation function $\phi$ and
the closely related energy-energy autocorrelation function.

\begin{figure}
\includegraphics[width=0.53\columnwidth,angle=-90]{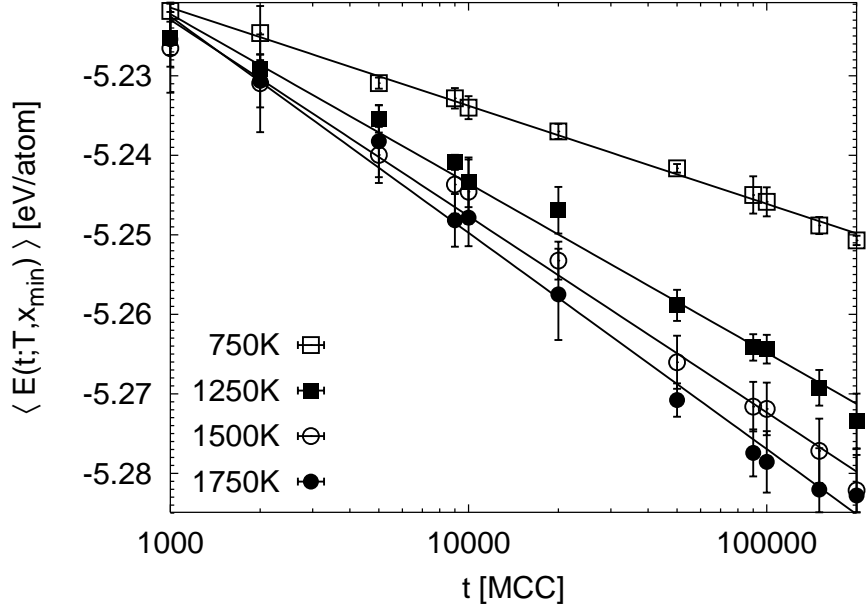}
\caption{Time dependence of the average energies
$\langle E(t;T,x_{min})\rangle = \langle E_{IS}(t,t_w = t_{init};T)\rangle $ of the minima $x_{min}$ for selected
temperatures $T=750$ K, $1250$ K, $1500$ K, $1750$ K as function of time $t= t_{init} + t_{obs}$.
\label{fig3}}
\end{figure}

The ensemble-averaged energy $\langle E_{IS}(t_w,t)\rangle$ of
the  inherent structures lying 'below' the trajectories
observed in the interval $[t_w,t_w+t_{obs}]$ decreases
logarithmically with $t_{obs}$, as  does the mean potential
energy $\langle
E(t_w,t_{obs})\rangle$. In both cases  the logarithmic slope
has a clear temperature dependence, but no significant
$t_w$ dependence for $t_w \ge t_{init} = 10^3$ MCC.
The energy simply decreases logarithmically with the simulation time elapsed after
the quench, i.e.\ to
a good approximation we find the functional form
\begin{equation}
E(t_{obs};T) = E_0(T) - A(T)\ln\left(\frac{t_{obs} + t_{init}}{t_{init}}\right).
\label{energy_drift}
\end{equation}
Here, $E_0(T) = \langle E(t_{init};T) \rangle$ and $E_0(T) = \langle E_{IS}(t_{init};T) \rangle$ for potential energies $E$ and minimum energies $E_{IS}$, respectively, and
\begin{equation}
A_{IS}(T) \approx 6.60\cdot10^{-6} \cdot T, \quad A_E \approx  6.64\cdot10^{-6} \cdot T,
\label{energy_slope}
\end{equation}
 where the coefficients are measured in eV/(K$\cdot$atom). Note that energies are reported on a
 per-atom basis.\newline
The logarithmic decay of the energies is shown in Fig.~\ref{fig3},
and has previously been observed  in other aging systems, see e.g.\ \cite{Kob00a,Kob00c,Angelani01a,Kob00b,Dall03}.

\begin{figure}
\includegraphics[width=0.53\columnwidth,angle=-90]{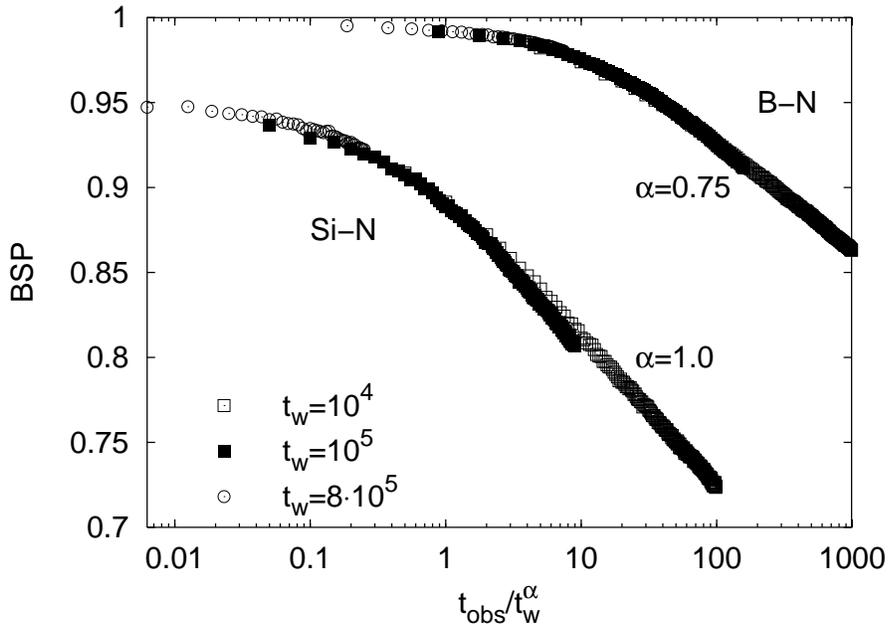}
\caption{Aging dependence  of the bond survival probability $BSP$ for Si-N and B-N bonds, respectively, as function of $t_{obs}/t_w^{\alpha}$, at temperature $T = 1250$ K.
\label{BSPaging}}
\end{figure}

Bond survival probabilities at $T=1250$ K for Si-N and B-N bonds are plotted  in Fig.~\ref{BSPaging} as a
function of the scaled variable $t/t_w^\alpha$ for
three different ages $t_w=10^4,10^5,8\cdot 10^5$ MCC. Even though the data collapse is not  perfect,
the scaling plots demonstrate that the bonds become, on average more resilient as the age increases.
For the Si-N bond, the best collapse is obtained
for  $\alpha_{SiN} =1$. Thus, the survival probability for the Si-N bond is
a function of $t/t_w$, a so-called pure aging behavior.
The survival probability for the B-N bond, however,
shows a sub-aging behavior with $\alpha_{BN} =0.75$.
For both types of bonds the survival probability decays
in a logarithmic fashion after a nearly flat initial part, and no cut-off is reached
at low temperatures, within the time scale of the simulations.\footnote{For temperatures different from 1250 K, the scaling exponents $\alpha$ can only be approximately determined due to the small ensemble of simulation runs at
these temperatures. The values for $\alpha$ lie in the range between 0.3 to 0.8}

Another quantity that exhibits an age dependence for $T < T_C$ is
the average mean square displacement of the atoms, given in Eq.~\ref{msd}.
We observe
a sub-diffusive behavior, $MSD(t_{obs},t_w) \propto t_{obs}^{\beta}$
with $\beta < 1$ ($\beta = 1$ corresponds to standard diffusion),
in the quasi-equilibrium region ($t_{obs} \ll t_w$).

\begin{figure}
\centering
\subfigure[MSD vs.\ $t_{obs}$]{
\includegraphics[width=0.33\textwidth,angle=-90]
{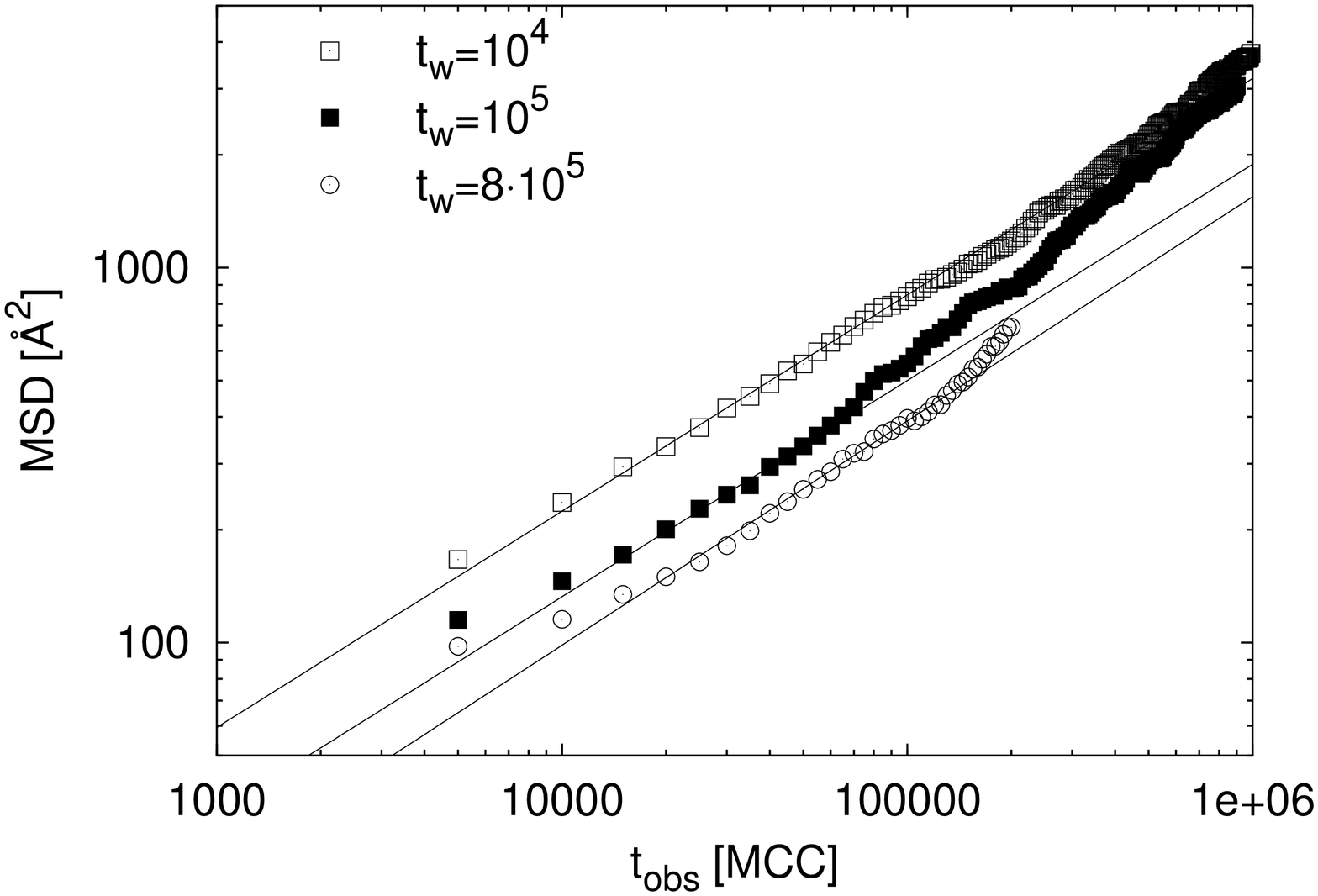}
\label{MSD_aging1}
}
\subfigure[MSD vs.\ $t_{obs}/t_w^{\alpha}$]{
\includegraphics[width=0.33\textwidth,angle=-90]
{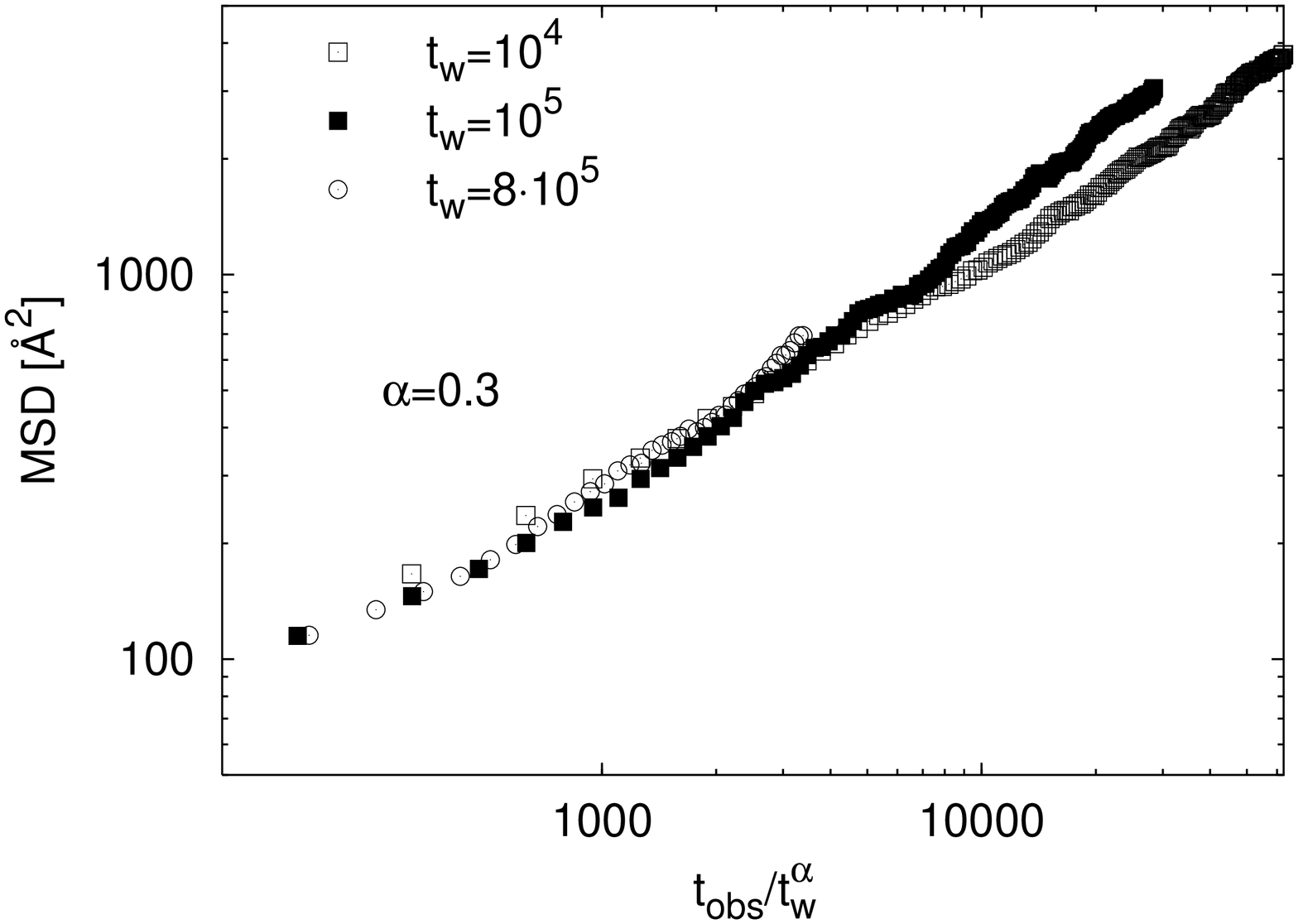}
\label{MSD_aging2}
}
\caption{Double logarithmic plot of the mean square displacement as function of observation time $t_{obs}$ for different
waiting times $t_w$, at temperature $T = 1250$ K. The exponent $\beta \approx 0.5$ of the sub-diffusive motion $MSD
\propto t_{obs}^{\beta}$ for waiting time $t_w$ is found by a fit of the data to a straight line, where only points up to
the waiting time $t_w$ were included. $\beta$ appears to be approximately independent of waiting time for $t_{obs} <
t_w$. Although the collapse of the curves after $t_w^{\alpha}$-scaling in subfigure b) is not perfect, due to the fact
that $\beta$ starts increasing for $t_{obs} > t_w$, the data indicate strong sub-aging to be present.
\label{MSD_aging}} \end{figure}

The $MSD$ remains sub-diffusive for $t_{obs} > t_w$
but with an increase in the exponent $\beta$.
In figure \ref{MSD_aging1}, we show the mean square displacement
 $MSD(t_{obs},t_w)$ for three different waiting times $t_w$,
 at temperature $T = 1250$ K. For $t_{obs} \ll t_w$,
 we find $\beta = 0.5$, and the curves can be
 approximately collapsed (c.f.\ figure \ref{MSD_aging2})
 if $t_{obs}$ is scaled with $t_w^{\alpha}$.
 The exponent $\alpha = 0.3$ indicates a strong  sub-aging
  behavior of the mean square displacement.
  This analysis was performed separately for
  all three types of atoms, Si, B and N.
  In all three cases, essentially the same results were found.

The two-point correlation function $\phi(t_w,t_{obs};T)$ (c.f.\ fig.~\ref{fig_phi}) remains very close to
the equilibrium value  $1$ in the ergodic temperature region ($T = 2500$ K), and does not depend on waiting time.
But in the glassy phase (T = 1750 K and T = 1250 K), $\phi(t_w,t_{obs};T)$ strongly deviates from its equilibrium value ($\phi
\equiv 1$) for $t_{obs} \ge t_w$. Quasi-equilibrium behavior is only observed for observation times $t_{obs} < t_w$, that
 grow monotonically with $t_w$.\footnote{The closely related autocorrelation function
$C_E(t_w,t_{obs};T) \equiv \langle
E(t_w) \cdot E(t_w+t_{obs})\rangle- \langle
E(t_w)\rangle\cdot\langle E(t_w+t_{obs})\rangle$
also exhibits the expected aging behavior, i.e. $C_E$ decreases to zero
from an almost  constant value ($\propto C_V(t_w)$) once
$t_{obs}$ exceeds $t_w$. Unfortunately, the data is too noisy
for us to draw any quantitative conclusions, beyond the
observation that the quasi-equilibrium regime grows monotonically with $t_w$.}

 This monotonic dependence on $t_w$ of the time range during
which quasi-equilibrium behavior is still observed, correlates with the
stiffening of the response of the system characteristic for aging processes:
The longer the system is allowed to equilibrate, the longer is the subsequent time range  during which
equilibrium-like behavior is observed. This effect concurs with the logarithmic
decay of $E(t;T)$ and   $\langle E(t;T)\rangle$ (c.f.\ figure \ref{fig3})
and fully agrees with  the general aging behavior previously  discussed.

\section{Discussion}
\label{Section_discussion}
Numerical investigations of atomistic models with
realistic interactions are computationally very demanding, and
for the same numerical effort,   the accuracy of the
results  obtained  is  lower than what can be achieved for
more simplified models. In spite of this limitation,
we have shown how ergodicity is broken in a-Si$_3$B$_3$N$_7$
and have characterized the dynamics in the low temperature phase,
where the model behaves as a bona-fide aging system.
In the aging regime the microscopic time scale of the dynamics is largely
irrelevant, and so is the precise value of the conversion factor between the time scales
of Monte Carlo and molecular dynamics simulations. This assumption is supported by the
fact that we have observed a similar slow drift in energy at low temperature MD simulations,
i.e.\ in the non-ergodic regime, during studies of heat transport in a-Si$_3$B$_3$N$_7$.\cite{Schoen02a} Therefore, we trust that the qualitative and most of the quantitative aspects of
our findings can be carried over to the real materials.

The computational analysis  of a-Si$_3$B$_3$N$_7$ using
MSD, BSP, $C_V^{a,b,c}$ and $\phi(t_w, t_{obs};T)$
 shows that this amorphous material can be expected to
 exhibit a glass transition with a concurrent break in
 the ergodicity at about $T_c \approx 2000$ K if
 pressures high enough to prevent decomposition are applied.\footnote{Repeating these investigations for a
 large number of densities,
we always find a similar freezing-in of the structure for $T\approx T_c$.
In other work, we have determined a critical point in the   liquid-gas region
of the ternary system ($p_{cr} \approx 1.3$ GPa and, $T_{cr}  \approx 8000 $K)
\cite{Hannemann03d}.
Since the tendency to decompose is greatly reduced for a supercritical fluid,
we predict that a-Si$_3$B$_3$N$_7$ should exhibit a
glass transition at a  temperature $T_G\approx 1700 - 2500 $ K and at a pressure of
$p_G > 2$ GPa.
Up to now, high-pressure experiments have only been performed for $T
\approx 1000$ K and $p \approx 2.5$ GPa.
}

For $T < T_c$, a-Si$_3$B$_3$N$_7$ exhibits a  general behavior
in many respects similar to standard test systems for computer simulations
~\cite{Sastry99a,Barrat99a,Crisanti00a} (Lennard-Jones, a-SiO$_2$),
insofar as we observe a freezing-in of the structure,
a logarithmic decay of the energy, and a systematic waiting-time dependence of the
bond survival probability, the mean square displacement of the particles and the two-time energy correlation
function.

The aging behavior  of the heat capacity  has  been studied experimentally,  for charge-density-wave
systems\cite{Biljakovic91a}, and for a spin-glass model~\cite{Sibani04}
and  discussed theoretically for  two-level systems at very low temperatures\cite{Parshin93a}.

The  logarithmic drift towards lower energies occurring for $T < T_c$  is also   observed in a number of model systems such as
Lennard-Jones \cite{Kob00c} and soft-sphere glasses\cite{Angelani01a}. This applies both to the
 actual trajectories and the time-sequence of observed local minima.
For fixed simulation time, the deepest local minima are reached for $T = 1750$
K, which lies right below $T_c$. Analogous observations are well-known from e.g.\
global optimization studies of complex systems, where it has been found that reaching the deepest
local minima using Monte-Carlo-type search algorithms is achieved by spending most of the search time
in the temperature interval slightly below the glass transition temperature\cite{Kirkpatrick83a,Salamon02a}.
Thus this result provides  another confirmation that ergodicity
breaking takes place at $T \approx T_c$.

In summary, all the numerical evidence obtained  from our
simulations shows that the thermal quench of a Si$_3$B$_3$N$_7$ melt produces a material
in a state of strong thermodynamic disequilibrium, which then relaxes
through the aging process. As we have noted in the introduction, a-Si$_3$B$_3$N$_7$ is currently only
 synthesized via the sol-gel route. In earlier work\cite{Hannemann04a}, we have shown that different
  synthesis routes (quench from melt, sol-gel, sintering of nano-crystallites, etc.) result in
  amorphous compounds that exhibit a monotonic decrease in energy during simulations at T $<$ 2000 K.
 This applies in particular to the relaxation of the structures that result from modelling the sol-gel synthesis route\cite{Schoen04a, Hannemann05a}, as we
   observed in both constant volume and constant pressure simulations \cite{Hannemann05b}. Thus,
   we surmise that aging phenomena analogous to those investigated here for a quenched melt also occur in
   the amorphous ceramic generated via the sol-gel route.

  The  metastable states
reached during aging become gradually lower in energy,  which could
lead to a strengthening
 of the material. On the other
 hand,  the structural transformations which
 would accompany aging could lead to the formation of  nano-crystallites
 and internal 'grain boundaries', whence
the overall structure might develop e.\ g.\ cracks, and thus lead to a weakening
of the
mechanical properties of the material.
This could especially be the case, if crystallites of
binary or ternary phases are formed.
However, for most envisaged applications, the time scales on which
such processes
 might occur are  beyond  the accessible
 simulation time by many orders of magnitude.

\section{Acknowledgments}
Funding was kindly provided by the DFG
via SFB408.


\end{document}